\documentclass[11pt]{article}

\usepackage{graphics}
\title{How Frustrated Strings Would Pull the Black Holes
from the Centers of Galaxies}

\author{Glenn D. Starkman  and Dejan Stojkovic \\
Dept. of Physics, Case Western Reserve University, Cleveland, OH 44106}

\begin{document}
\maketitle

\newcommand{\be}{\begin{equation}}
\newcommand{\ee}{\end{equation}}
\newcommand{\dy}{\dot{y}}
\newcommand{\half}{\frac{1}{2}}
\newcommand{\sun}{{\rm Sun}}

\abstract{Recently Pen  and Spergel (1997) have shown that  a universe
whose energy density is dominated by a frustrated network of
non-Abelian TeV-scale cosmic strings could account for a broad
class of cosmological observations.
In this paper we consider the effects of such a string network
on the massive black holes widely believed to inhabit the centers
of many galaxies.  As these black holes traverse the universe
together with their host galaxies, they would intersect
a large number of string segments.  We argue that such segments
would become stuck to the black hole, and be stretched by the
hole's motion.  Stretching the strings would cause significant
deceleration of the black holes.  Although the black holes
would probably not be removed from the galaxies completely,
they would be  noticeably displaced from the galactic center of mass --
by at least 5kpc.  This displacement seems to be is contradiction to the
observational evidence.

\section{Introduction}

Recently, Pen and Spergel (1997) proposed that
the energy density of the universe is dominated
by the contribution from a frustrated network of non-Abelian
cosmic strings. Because the network is frustrated, the
string energy  density scales as only the inverse square of the scale factor.
This results in a non-decelerating  (though also non-accelerating)
universe,  and so an alternative cosmology intermediate between
dark matter dominated Open CDM and cosmological-constant dominated
$\Lambda$ CDM. \cite{S} presents all the basic features of the
frustrated string dominated models. This model is also discussed in
\cite{BS}.

The basis of Pen and Spergel's proposal  is that in some
non-Abelian field theories, there are cosmic strings which do not intercommute,
{\it i.e.} pass through each other, effectively.
Consequently, the string network can become frustrated.
It then ceases to effectively relax to its minimum energy configuration,
it ``freezes out."
The vacuum energy density of the strings can then
come to dominate the energy density of the universe.

Pen and Spergel refined this scenario with numerical simulations
of the evolution of a network of non-Abelian cosmic strings.
These simulations showed that for $\Omega_0
\sim 0.4-0.6$ and $H_0 \sim 60-70$ km/s/Mpc their model is consistent
with the current observations, including CMB fluctuations,
the shape of the galaxy power spectrum
the amplitude of the mass power spectrum, and limits on the age of the
galaxies. It was also consistent
with observations of high redshift supernova
and gravitational lensing statistics.

If the universe today is string dominated, the
strings must have formed at an energy scale near the
electroweak scale ($\sim 1$ TeV), implying an energy per
unit length of approximately $5\times10^{-5}$g/cm.
The characteristic comoving separation between   strings
is approximately $10^{-3}$ of
the bubble size during the phase transition,
corresponding to $L\simeq0.1$ A.U. today.
(An A.U.  is the distance between the earth and the sun.)

This density of strings  is surprising --
there would be many strings inside our Solar System.
However, because the mass per unit length of each string is so low,
they would be very difficult to detect.
For example the gravitational bending angle due to the string
is only $(M_W/M_{Pl})^2 \sim 10^{-32}$ radians,
where $M_W$ is the electroweak symmetry breaking scale
and $M_{Pl}$ is the Planck scale.
Even if the string can catalyze baryon decay,
its cross section is so small that the string could pass
through the sun, the earth, or even a person, completely undetected.
The probability that a string would pass through a specially designed
detector in a reasonable time is extremely low.

There is an exception to the innocuousness of the electro-weak
scale strings. If a black hole encounters a string at small
enough impact parameter that part of the string enters the event
horizon, then, as we will show, the black hole will capture the
string. If the black hole horizon is large compared to the string
cross-section, then the string will remain threaded through the
black hole horizon. The exterior ends are unlikely to reconnect
for an enormously long time.

Let us assume that such a network of ``frustrated strings"
really exists in our universe. We will show by simple arguments
that such a dense string network would severely impact
the motion of black holes in the centers of galaxies.

There is convincing evidence that almost all galaxies contain
very massive compact objects, probably black holes, at their centers.
Analysis of data such as galactic rotation curves,
luminosity and mass-to-light ratio indicates
the existence of a central extreme mass concentration in many galaxies,
including our Milky Way\cite{Ghez}, M87 \cite{M87} ,
and others \cite{3377} - \cite{4486B}.
The most convincing case for a black hole {\it per se}
is from our own Galaxy\cite{Ghez},
for which the stellar rotation curves about Sag A$^\star$
have been measured to approximately $0.01-0.03$pc,
and preclude the presence of a stable cluster of stars,
white dwarfs, neutron stars, or even stellar-mass black holes.
Other than a massive ($2.6\times 10^6M_\sun$) black hole
the only possible explanations are a super-dense cluster of either
very light ($<0.02M_\sun$) black holes or weakly interacting dark matter.
If anything, each of those seems more far-fetched
than a single massive black hole.

The range in mass of these black holes is $10^6 -
\mathrm{few}\times10^9M_\sun$, implying a range in Schwarzschild
radii of $10^9-\mathrm{few}\times10^{12}$m. Especially for the
more massive black holes, this is larger than the  inter-string
separation. Thus  as the black hole moved through the universe it
would encounter many strands of cosmic string. These strands
would be stretched by the  motion of the black  hole through
space. As the strings are stretched, they reduce the kinetic
energy of the black hole, slowing it.  Although each string has a
minimal effect on the black hole, even integrated over the
lifetime of the universe, the cumulative effect of the large
number of strings that are being stretched, can be very
significant. Typically we find that, despite the dynamical
response of the host galaxy, a typical galactic core black  hole
would be observably shifted from the center of the galaxy.

As we argued briefly above,
the  motions of ordinary stars are not likely to be affected --
the strings pass right through even a neutron star.
Stellar mass  black holes will be affected, albeit
to a lesser extent than galactic black holes.
This is because the cross-sectional area of a black hole,
and hence the  number of strings it is expected to capture,
is proportional to the square of the black hole mass.
The total black hole deceleration due to string drag is therefore
proportional to the first power of the black hole mass.

\section{String Network and a Black Hole}

Consider a super-massive black hole traversing the universe and
encountering a string. (We choose to view things in the rest
frame of the string, because the string network is frustrated and
nearly static in the frame of the CMB.)  Since the area of the
black hole is large compared to the string cross section, after
the black hole hits the string, the string sticks to the black
hole \cite{F}. The portion of the string which enters the horizon
cannot be pulled back out. The response of the rest of the string
is limited by causality, i.e. only the portion of the string
within the sound horizon $c_st$ of the  initial point of
string-black hole impact can respond to the encounter of the
string with the  black hole. Here $c_s$ is the sound velocity in
the string  (typically $c_s\simeq c$), and $t$ is the time
elapsed since the black-hole encountered the string. With the string fixed
beyond $c_st$ and the middle pinned to the black hole, the string begins to stretch.
The notion of the string length in the background space of the
black hole is different from the flat space one. However, by minimizing
the length of the string from some point outside the horizon to the
horizon itself, one can infer that, for a string velocity perpendicular
to the string, after the equator of the black hole passes the string,
the string prefers energetically to enter the black hole horizon at
normal incident, and the string thereafter stretches rather than recombining
behind the black hole.

An isolated string has a constant energy per unit length, and so
the string configuration which minimizes the length is also the
configuration which minimizes the total energy. Actually, there
is no just one string, but rather a string network. If the strings
were connected in an isolated two-dimensional surface, then the
minimum energy configuration would be the surface of  minimum
area. Since the frustrated non-Abelian strings are connected in a
highly three-dimensional network, the minimum energy configuration
is the configuration which minimizes the volume of the deformed
network.  By axial symmetry, the volume of the network is the
volume of rotation  of a curve $f(y)$ about the axis between the
original point of contact of the black hole with the network
($y=a$), and the black hole's position at some time $t$ later
($y=b$). The volume of rotation of a  curve $f(y)$ about the y
axis for $a<y<b$ is: \be V=\pi \int_a^b [f(y')]^2 dy' \ee Our
variational equation is: \be \delta \int_{0}^{y} [f(y')]^2 dy' =0
\ee \noindent with $f(0)=L$ and $f(y) = c_s t$. The solution is a
curve which is zero everywhere except at the end points -- the
configuration shown in  Fig. (\ref{fig1}). Note that string network can be
treated as a continuous medium only on scales bigger than the
characteristic distance, $L$, between the the string segments.
Thus, the width of the tube in Fig. (\ref{fig1}) is  approximately
$L$.

\begin{figure}[!ht]
\centering
\includegraphics{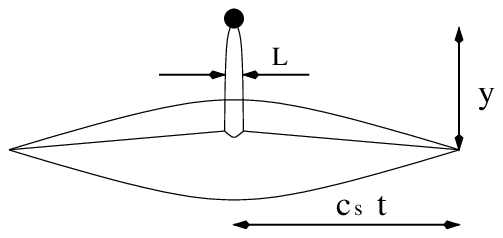}
\caption{\label{fig1}Actual shape of the deformed string medium}
\end{figure}

It should be noted that the response of a particular frustrated
string network configuration might be intermediate between the
volume-minimization and area-minimization  results quoted above
(or even the length minimization result). However, we will adopt
the most conservative possibility --- the minimization of volume
--- to obtain limits on the string network.

\section{String Recombination and Snapping}

All the effects which can result in a black hole shedding the
strings have to be taken into account. For example, if two ends
of (not necessarily the same) string on the black hole horizon
come within a distance characterized by the string thickness
($\sim\frac{1}{TeV}$) apart, then they can recombine and detach
from the black hole.

\begin{figure}[!ht]
\centering
\includegraphics{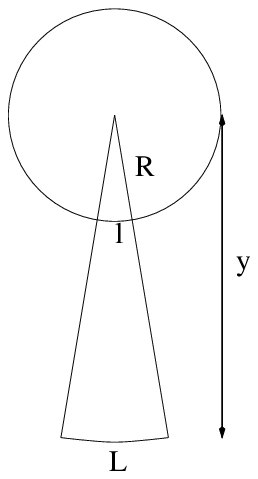}
\caption{String recombination}
\end{figure}

In traveling a distance $y$ through the universe,
the black hole accumulates $N(y) =\frac{y}{L} \frac{\pi R_{Sch}^2}{L^2 g}$ strings.
($g$ is  a geometrical factor having to do with the connectivity of the network.)
The ends of these strings will accumulate on a circle of radius
$l=\frac{R_{Sch}L}{y}$.   Recombination will happen when
$\frac{2\pi l}{N(y)} < \frac{1}{TeV}$, i.e. when the black hole has traveled
\be
y_{rec} \simeq \sqrt{\frac{2gL^4TeV}{R_{Sch}}}
\ee
For a $10^9$ Solar masses black hole,
and $g=3$  we get $y_{rec} \sim 2\times10^{7}$pc.
A black hole traveling at $(1-2)\times10^{-3}$c
would take ${\cal O}(10^{11})$ years to travel this distance.

One thing which could affect above calculation is string oscillations.
Collision of a string with a black hole would excite oscillatory modes
of the string. These oscillations would enhance the probability for
string segments to collide with each other.  However,   the
initial amplitude of these oscillations are relatively small
because the black hole is moving very slowly compared to the
sound speed in the strings.  Also, these
string oscillations are rapidly damped by several effects.
First, the energy of string oscillations are rapidly propagated away
from the black hole into the bulk network.  Second, the
links nearest the black hole are being stretched, effectively
``red-shifting'' away the oscillations.
Also, for most non-Abelian string networks, most string-string collisions
do not result in reconnections which result in a string loop falling
into the black hole.  Indeed, because generic reconnections would be with
string links that were captured by the black hole long before,
they would cause the network to become more tangled, rather than
less, and hence not result in any decrease in the force
applied by the stretching string network on the black hole.
Correct and full treatment of these
oscillations requires a concrete model of the strings and
probably a detailed numerical evolution of the
frustrated string network in the black hole wake.

There is also possibility that the strings might snap. The
snapped ends could terminate on monopoles or mini black holes and
the black hole could tear the network rather than get caught.
However, one can easily show that the probability for this to
happen is negligible. If the underlying field theory contains
monopoles, then the probability of monopole-antimonopole
nucleation per unit length per unit time is $\sim M_W^2 e^{-\pi
m_m^2/u^2}$ where $m_m$ is the mass of the monopole and $u^2$ is
the string tension or the energy per unit length of the string
\cite{Vil}. From $m_m \sim 4\pi M_m/\alpha$, where $\alpha$ is
the fine structure constant and $M_m$ is the energy scale at
which the monopoles were formed, we see that the process is
suppressed by a huge exponential factor even if $M_m \sim M_W$.
Even integrated over the lifetime of the universe and over the
causal length corresponding to this time, the probability is
still negligibly small due to the non-perturbative nature of this
effect. The snapped string ends could also terminate on mini
black holes. For a mini black hole of the Planck mass the
probability of pair nucleation per unit length per unit time is
$\sim M_{W}^2 e^{-\pi M_{Pl}^2/u^2}$ which is even more
suppressed than the monopole-antimonopole creation. Finally, in
the case when the underlying field theory does not support
monopoles, the presence of the black hole changes the topology of
the space-time and allows monopole solutions which wind around the
black hole or the black hole itself could carry magnetic charge
and be a monopole itself. Due to their enormous mass, the
probability of nucleating of such monopoles is enormously small.

\section{Deviation from the Free-Motion Path due to Strings}

The change in energy of one string due to the motion of the black hole
a distance $\Delta y$ past their initial point of  encounter
is $\Delta E \approx 2 \Delta y u^2$, where $u^2$ is energy per
unit length of the string. The total change in energy due to all the strings
that the black hole encounters is
\be
(\Delta E)_{tot}=\frac{\pi R_{Sch}^2}{L^3 g} u^2 \int_{0}^{y} 2 ( y-y_0)
dy_0 = \frac{\pi R_{Sch}^2 u^2 y^2}{L^3 g}
\ee

Imposing energy conservation we find:

\be \label{ee}
\half M_{bh} \dy _0^2  =
\half M_{bh} \dy^2 + \frac{\pi R_{Sch}^2 u^2 y^2}{L^3 g}
\ee
\noindent where $\dy _0$ is the initial velocity of the black hole.

Let $\Delta = \dy_0 t - y$ measure the deviation of the black hole from its
free path. Suppose that $\Delta$ is small, i.e.
$\Delta \ll \Delta_{max} \equiv \dy_0 t$.
Linearizing in $\Delta$ we find:
\be \label{lin}
\dot{\Delta} = \frac{\pi R_{Sch}^2 u^2 \dy_0}{g L^3 M_{bh}}\ t^2 ,
\ee
\noindent the solution of which is
\be \label{sol}
\Delta = \frac{\pi R_{Sch}^2 u^2 \dy_0}{3 g L^3 M_{bh}}\ t^3  .
\ee

(Note that equation (\ref{ee}) can be solved exactly giving
\be
y= \dot{y_0} \sqrt{\frac{g L^3 M_{bh}}{2 \pi R_{Sch}^2 u^2}}
\sin \left( \sqrt{\frac{2 \pi R_{Sch}^2 u^2}{g L^3 M_{bh}}} t \right)
\ee
and
\be
\Delta=\dot{y_0} \left[ t -  \sqrt{\frac{g L^3 M_{bh}}{2 \pi R_{Sch}^2 u^2}}
\sin \left( \sqrt{\frac{2 \pi R_{Sch}^2 u^2}{g L^3 M_{bh}}} t \right) \right]
\ee
This is a monotonicly increasing function which for a small argument expansion
in $\sin$ gives the result (\ref{sol}).)

Taking $u^2 =(1 TeV)^2$, $\dy_0=300$km/s,
$g=3$, $L=0.1$ A.U., $M_{bh}=10^9M_{sun}$ and the elapsed
time since the black hole began moving to be only $t=3 \cdot 10^9$ years,
we get:
\be \label{res}
\Delta \sim 3 \cdot 10^6 \mathrm{pc}
\ee

Obviously, $\Delta $ is bigger than $\dy_0 t \sim 3 \cdot 10^5$pc
which means that our assumption $\Delta \ll \Delta_{max} \equiv \dy_0 t $
would break down.
However, result (\ref{res}) has ignored the gravitational pull of the galaxy
in which the black hole is embedded, which will counteract  the
deceleration due to the  string.

\section{Influence of the Galaxy}

Because the force on the black hole due to the stretching of the string network
is quite small,   the displacement of the black hole relative to the center of
its host galaxy is slow enough that the galaxy can respond dynamically.
We must therefore include the gravitational  pull of the galaxy on the black hole
in the black hole equations of motion.
Looking from the rest galaxy frame,
the displaced black hole will start to drag a part of galaxy with it.
This could be an considerable effect -- if the galaxy were a rigid body
attached to the black hole, then the acceleration of the black hole would
be reduced by a factor of $M_{Galaxy}/M_{bh} > 10^3$.
The galaxy is not a rigid body, nor is it rigidly attached to the black hole,
hence we must model the galactic mass density distribution, and the
galaxy response, although the
details of the model will have little effect on our final conclusions.
We adopt the following simple, but conservative,  model --
we take the galaxy to be spherically symmetric
with the mass density excluding the black hole given by
\be
\rho (r) = \rho_0 \left\{ \begin{array}{lll} 0  & r < r_c \\
(\frac{r}{r_0})^{-(3+\gamma)}  &  r_c < r < r_0 \\
(\frac{r}{r_0})^{-2}  &   r > r_0
\end{array} \right.
\ee
The region $r<r_c \equiv R_{Sch}$ is occupied by the black hole.
The region $r_c < r < r_0$, where $\rho (r) \sim r^{-3}$ or steeper,
is the central core of the galaxy. Empirically, $r_0\simeq2-10 \mathrm{kpc}$.
In the third region ($r>r_0$), $\rho(r)$ goes more or less like $r^{-2}$.
$\rho_0$, the density at the radius where the velocity dispersion or
orbital velocity turns over, can be determined from the condition
$\frac{G (M(r_0) + M_{bh})}{r_0^2} = \frac{v_0^2}{r_0}$, where $v_0$ is the
linear orbital velocity of galaxy at the distance $r_0$ from the center of
the galaxy.
Accordingly, the galactical mass distribution
$ M(r) \equiv 4 \pi \int_{0}^{r} \rho (r) r^2 dr $ is:
\be
M(r) = \left\{ \begin{array}{ll} \frac{(\frac{v_0^2 r_0}{G} - M_{bh})}
{(1-(\frac{r_c}{r_0})^{\gamma})} (1-(\frac{r_c}{r})^{\gamma}) &
\gamma >0, \ r_c<r<r_0, \\
\frac{(\frac{v_0^2 r_0}{G} - M_{bh})}{ln(\frac{r_0}{r_c})}
ln(\frac{r}{r_c}) & \gamma = 0, \ r_c <r<r_0 .
\end{array} \right.
\ee
We then assume that if the black hole has been displaced by
the string-force by $\Delta$ from the center of the galaxy,
then it carries with it that portion $M(\Delta)$ of the galactic
mass within the radius $\Delta$.

The energy conservation equation (\ref{ee}) is now replaced by:

\be \label{gee}
\half M(\Delta) \dy_0^2 + \half M_{bh} \dy_0^2 =
\half M(\Delta) \dy^2 + \half M_{bh} \dy^2 + \frac{\pi R_{Sch}^2 u^2 y^2}{L^3 g}
\ee
\noindent Again, we are interested in small $\Delta$, obviously much smaller
than $\dy_0 t$, but we also have much stronger constraints.
If $\Delta$ is of the order of $r_0 \sim 2-10 \mathrm{kpc}$,
then it means that the black hole has been displaced $2-10 \mathrm{kpc}$
from the center of the galaxy -- clearly inconsistent with the
observations of elliptical galaxies with candidate black holes.
We therefore take $\Delta < r_0$. Linearizing again in $\Delta$,
(\ref{gee}) can be rewritten:
\be \label{eq}
 \dot{\Delta} = (M(\Delta) + M_{bh})^{-1}\frac{\pi R_{Sch}^2 u^2 \dy_0}{L^3 g} t^2  .
\ee
Since $M(\Delta) =0$ for $\Delta < r_c$,
in this region $\Delta$ is given by equation (\ref{sol}).
The black hole therefore reaches $\Delta=r_c$ in time
\be
t_c=(\frac{3 g L^3 M_{bh} r_c}{\pi R_{Sch}^2 u^2 \dy_0})^{\frac{1}{3}}
\ee
For $t>t_c$ (or equivalently $\Delta > r_c$) we have two cases,
$\gamma = 0$ and $\gamma =1$. Integration from $t_c$ to $t$ gives:
\be\label{g0}
\frac{(\frac{v_0^2 r_0}{G} - M_{bh})}{ln(\frac{r_0}{r_c})}
(ln(\frac{\Delta}{r_c}) -1) \Delta + M_{bh} \Delta =
\frac{\pi R_{Sch}^2 u^2 \dy_0}{L^3 g} \frac{t^3}{3} +
\frac{\frac{v_0^2 r_0}{G} - M_{bh}}{ln(\frac{r_0}{r_c})} r_c
\ee
when $\gamma=0$, and
\be\label{g1}
\frac{(\frac{v_0^2 r_0}{G} - M_{bh})}{(1-\frac{r_c}{r_0})}
r_c \, ln\frac{\Delta}{r_c} +
(\frac{(\frac{v_0^2 r_0}{G}-M_{bh})}{(1-\frac{r_c}{r_0})} + M_{bh}) \Delta =
\frac{\pi R_{Sch}^2 u^2 \dy_0}{L^3 g} \frac{t^3}{3} +
\frac{(\frac{v_0^2 r_0}{G} - M_{bh})}{(1-\frac{r_c}{r_0})} r_c
\ee
when $\gamma>0$.
Using $M_{bh}=10^9M_\sun$, $\Delta_{max}=r_0 \sim 3$kpc,
$v_0 = 300$km/s,  $t=3\times10^9$ years and $(ln \frac{\Delta}{r_c})_{max} \sim 17$,
we find $\Delta \sim 50$kpc for both $\gamma =0$ and $\gamma =1$.

This result should not, or course, be believed for $\Delta > r_0$, but even
$\Delta = r_0$ is enough to rule out the frustrated string network model.
The time needed to reach $\Delta=r_0$.
is $t_0 \sim 10^9$ years for both $\gamma =0$ and $\gamma =1$.
We can also estimate the displacement $\Delta$ of the black hole
in the last $\Delta t = 250 \cdot 10^6$ years, which is a characteristic
dynamical time of the galaxy.
For this purpose, we integrate equation (\ref{eq})
from ($t_{now}- \Delta t$) to $t_{now}$ with the very conservative assumption that
$\Delta(t_{now}- \Delta t)=0$.
For both $\gamma=0$ and $\gamma=1$, we find $\Delta \sim 5$kpc.

If we accept that the astronomical observations indicate
massive black holes, some of mass greater than $10^9M_\sun$,
located within 400pc of the galactic center (400pc is an underestimate of
the accuracy of the black hole position \cite{BKS}),
then the possibility of a cosmologically dominant
frustrated non-Abelian string network is severely complicated.

\section{Conclusion}

The possibility that a frustrated network of non-Abelian strings
dominates the energy density of the universe is a fascinating one.
Although there are models of such strings which
agree remarkably well with a broad class of observations,
we have shown that with the string and network parameters required
to play the desired cosmological role,
the  string network would severely affect the relative motion of
central galactic black holes with respect to the rest of the galaxy,
if the black hole mass is greater than about $10^8M_\sun$.
The black hole would be noticeably displaced from the center
of the galaxy even in the last galactic dynamical time.
This does not necessarily rule out a cosmologically important
frustrated non-Abelian string network.  It could for example be that
the superdense objects at the centers of galaxy are not black holes,
but some other exotic objects which would interact far more weakly
with the string network. One might also be able, by altering
the field-theory model which gave rise to the strings,
to alter the strings' linear density or network properties
and thereby reduce the dynamical effects on the galactic black holes.
In any case, these effects must be considered in any ongoing attempts
to incorporate cosmic string networks into cosmology.

\end{document}